\documentclass[aps,pra,reprint,showpacs]{revtex4-1} %groupedaddress
\usepackage{graphicx}  % needed for figures
\usepackage{dcolumn}   % needed for some tables
\usepackage{bm}
\usepackage{verbatim}   % for math
\usepackage[version=3]{mhchem} % Package for chemical equation typesetting
\usepackage{amsmath, amsfonts, amssymb} % math equations, symbols
\usepackage{textcomp} % This package is just to give the text quote '
\usepackage{enumitem}
\usepackage{color}
\usepackage{psfrag}
\usepackage{subfigure}
\usepackage{physics}
\usepackage{subfigure}
\usepackage{float}
\usepackage[raggedright]{titlesec}
\usepackage{titlesec}

\begin{document}

% full title (Capitalized)
\title{Shannon Entropy and Diffusion Coefficient in Parity-Time Symmetric Quantum Walks}

% authors (add full first names)
\author{Zhiyu Tian $^{1,\dagger}$, Yang Liu $^{1,2,\dagger}$ and Le Luo $^{1,2,\ast}$}

% affiliations/addresses
\affiliation{
$^{1}$ School Of Physics And Astronomy, Sun Yat-Sen University, Zhuhai, 519082, China \\
$^{2}$ Center Of Quantum Information Technology, Shenzhen Research Institute Of Sun Yat-Sen University, Nanshan Shenzhen 518087, China\\
$^{\dagger}$ These authors contribute equally to this work\\
$^{\ast}$ Correspondence: luole5@mail.sysu.edu.cn}
%{\em Received: ; in revised form: / Accepted: / Published: }}
%\end{abstract}
% the abstract

\begin{abstract}
Non-Hermitian topological edge states have many intriguing
properties, but have so far mainly been discussed in terms of
bulk-boundary correspondence. Here we propose to use a bulk property
of diffusion coefficients for probing the topological states and
exploring their dynamics. The diffusion coefficient is found to show
unique features with the topological phase transitions driven by
parity-time($PT$)-symmetric non-Hermitian discrete-time
quantum walks as well as by Hermitian ones, despite artificial
boundaries are not constructed by inhomogeneous quantum walk. For a
Hermitian system, a turning point and abrupt change appears in the
diffusion coefficient when the system is approaching the topological
phase transition, while it remains stable in the trivial topological
state. For a non-Hermitian system, except for the feature associated
to the topological transition, the diffusion coefficient in the
PT-symmetric-broken phase demonstrates an abrupt change with a peak
structure. In addition, the Shannon entropy of the quantum walk is
found to exhibit a direct correlation with the diffusion
coefficient. The numerical results presented here may open up a new
avenue for studying the topological state in Non-Hermitian quantum
walk systems.
\end{abstract}
\maketitle

% add 3 or more keywords
%\keywords{Shannon entropy; diffusion coefficient; quantum walk.}

%%%%%%%%%%%%%%%%%%%%%%%%%%%%%%%%%%%%%%%%%%%%%%%%%%%%%
\section{Introduction}

Quantum walk
\cite{childs2009universal,karski2009quantum,schmitz2009quantum,lovett2010universal,zahringer2010realization,childs2013universal}
is the quantum analog of the classic random walk, which has found
wide applications in many realm of quantum information science. In
recent years, quantum walks have been extensively studied
theoretically, and their experimental realizations have been
reported by single neutral atoms in optical lattices
\cite{karski2009quantum}, photons in waveguide lattices
\cite{perets2008realization}, trapped ions
\cite{schmitz2009quantum,zahringer2010realization} and single
photons in free space \cite{broome2010discrete}. Since quantum walk
can be constructed with various symmetries by designing appropriate
evolution operators with periodic driving, it enables readily
observation of the topological phenomena with Floquet methods.
quantum walk has been proved to be suitable for simulating
topological materials and exploring dynamics in a wide
range\cite{thouless1982quantized}

Recently, the use of quantum walks to simulate and
explore new topological
phenomena\cite{kitagawa2010exploring,kitagawa2012observation,cardano2017detection}
has attracted wide interests. Particularly, a new type of quantum
walk, open quantum walks\cite{attal2012open,attal2012open2} have
been realized and investigated, where the topological properties can
be controlled by dissipation and decoherence. Open quantum walks
show rich dynamical behavior
\cite{attal2012open,attal2012open2,sinayskiy2013open,bauer2013open}
and can be used for quantum algorithms in quantum computation for
specific tasks\cite{sinayskiy2012efficiency} and quantum state
engineering\cite{attal2012open,sinayskiy2012efficiency}.
PT-symmetric quantum walks, as one type of open quantum walks, have
recently been investigated both theoretically
\cite{mochizuki2016explicit,mittal2021persistence} and
experimentally \cite{xiao2017observation,dadras2018quantum}.

However, physical implementation of PT-symmetric
quantum walks have been limited due to a few issues. First is the
limited evolution time induced by dissipation and decoherence
effects\cite{ryu2002topological}. These effects are unavoidable in
open quantum walks\cite{attal2012open,attal2012open2}, which is
because of the non-unitary dynamics and is described in the
framework of the non-Hermitian quantum mechanics
\cite{bender1998real,bender2002complex}. Second, PT-symmetric
quantum walks and related topological properties have only been
studied by the bulk-boundary correspondence
\cite{yao2018edge,kunst2018biorthogonal,mochizuki2020bulk}, while
the characterization of bulk property is absent so far, such as
diffusion property and Shannon entropy. Until very recently, the
connection between diffusion property and topological property have
been studied for skyrmions \cite{weissenhofer2020diffusion}, and the
relation between diffusion phenomenon and bulk-edge correspondence
has been discussed \cite{yoshida2021bulk}. 

In this article, we propose to use diffusion
coefficient, one of the bulk properties, for characterizing the
topological phases in PT-symmetric quantum walk. With this new
viewpoint, we can address the above-mentioned two issues and there
is no necessity of constructing artificial boundaries. We further
illustrate the direct correlation between Shannon entropy and
diffusion coefficient for the first time. The result suggests a
fascinating possibility for exploration topological properties of
non-Hermitian systems
\cite{zeuner2015observation,weimann2017topologically} using
diffusion coefficient. It is noted that our main tool is the
numerical simulation instead of the analytical method so that the
general principle can be further revealed by analytical calculation.

\section{Discrete-Time Quantum Walk}
The quantum walk is governed by the time-evolution operator \cite{xiao2017observation}
\begin{equation}
U=\operatorname{LTC}\left(\beta\right) L^{\prime} T C\left(\alpha\right)
\end{equation}
where $T$ is the position operator
\begin{equation}
T=\sum_{x}|x+1\rangle\langle x|\otimes| \uparrow\rangle\langle\uparrow|+| x-1\rangle\langle x|\otimes| \downarrow\rangle\langle\downarrow|
\end{equation}
where x is the position of the walker. The position state changes according to the coin states $| \uparrow\rangle$
and $| \downarrow\rangle$ of the walker. $C$ is the position-dependent coin operator:
\begin{equation}
C(\theta)=I \otimes\left(\begin{array}{cc}
\cos \theta & \sin \theta \\
\sin \theta & -\cos \theta
\end{array}\right)
\end{equation}
where $I$ is a unit matrix with the same dimension as the number of grid points. The loss operators
are defined as:
\begin{equation}
L=I \otimes\left[\begin{array}{ll}
l_{1} & 0 \\
0 & l_{2}
\end{array}\right], L^{\prime}=I \otimes\left[\begin{array}{ll}
l_{2} & 0 \\
0 & l_{1}
\end{array}\right]
\end{equation}

The evolution operator of the discrete-time quantum walk can be written in the form of $U=e^{-i H_{eff}}$, where $H_{eff}$ represents the equivalent Hamiltonian of the system. Let $\lambda$ as the eigenvalue
of the evolution operator, then we have
\begin{equation}
U|\psi\rangle=\lambda|\psi\rangle, \lambda=e^{-i \epsilon+\epsilon_{0}}
\end{equation}
where $\epsilon$ is the quasi-energy of the system $e^{-\epsilon_{0}}=1 / l_{1} l_{2}$. When $l_1=l_2=1$ the evolution operator $U$ is a unitary matrix with $\epsilon_0=0$ and $\abs{\lambda}=1$. The system is a Hermitian system. When $l_1\ne l_2$, it is a non-Hermitian system and $U$ is a non-unitary operator, which respects PT-symmetry $(\mathcal{P T}) U(\mathcal{P T})^{-1}=U^{-1}$.

\section{Diffusion Coefficient and Shannon Entropy}

As demonstrated in \cite{xiao2017observation}, the topological phase of the evolution operator can be changed by
tuning two variables of the system evolution operator $\alpha$ and $\beta$. There are boundaries between
topological phases with different topological invariants. When $\left(l_{1}, l_{2}\right)=(1,1)$ which corresponds
to no loss, the evolution operator is the Hermitian operator. In this case, the topological invariant
at the boundary is not defined. When $\left(l_{1}, l_{2}\right)=(1,0.8)$, the evolution operator is a
non-Hermitian operator, and there is a boundary area between different topological phases, where
the PT-symmetry of the state is broken, and the topological invariant is not defined. It has
been proved that whether Hermitian operator or PT-symmetric non-Hermitian operator is used to
construct an artificial boundary in the system, there will always be a topological edge state at the
boundary when evolution operators with different topological invariants are used on each side of the boundary \cite{xiao2017observation}, which is related to the bulk-boundary correspondence in the topological effects
\cite{essin2011bulk}.

In order to study the topological effect of the quantum walk, traditionally, artificial boundaries
are constructed by employing different evolution operators on each side of
the boundary, then the probability distribution at the boundary are probed, thus enabling
simulation of the bulk-boundary correspondence in topological materials \cite{kitagawa2010exploring}. Recently, time evolution of the position variance \cite{brun2003quantum, romanelli2005decoherence, xue2014trapping}, statistical moments of the walker position distribution \cite{cardano2016statistical, wang2018detecting, xiao2018higher}, are measured for either investigating the localization effects or revealing the topological phase transitions in discrete-time photonic quantum walks.

Here we adopt a different approach, that is using the diffusion coefficient \cite{romanelli2005decoherence,zhang2017topology}, a natural feature of the quantum walk
as the probe of the topological edge states, since the evolution of the system during the quantum
walk would lead to a distinctive diffusive behavior \cite{romanelli2005decoherence}. In addition, we investigate the
properties of Shannon entropy in the quantum walk. In the following sections, we simulate the
quantum walk, both Hermitian and PT-symmetric non-Hermitian, and observe the behavior of
the diffusion coefficient and Shannon entropy. In all cases we take the initial state as $|\psi\rangle=|0\rangle$.

\subsection{Hermitian Quantum Walk}

\begin{figure*}
\centering
\subfigure{
\begin{minipage}[t]{0.4\linewidth}
\centering
\includegraphics[width=3in]{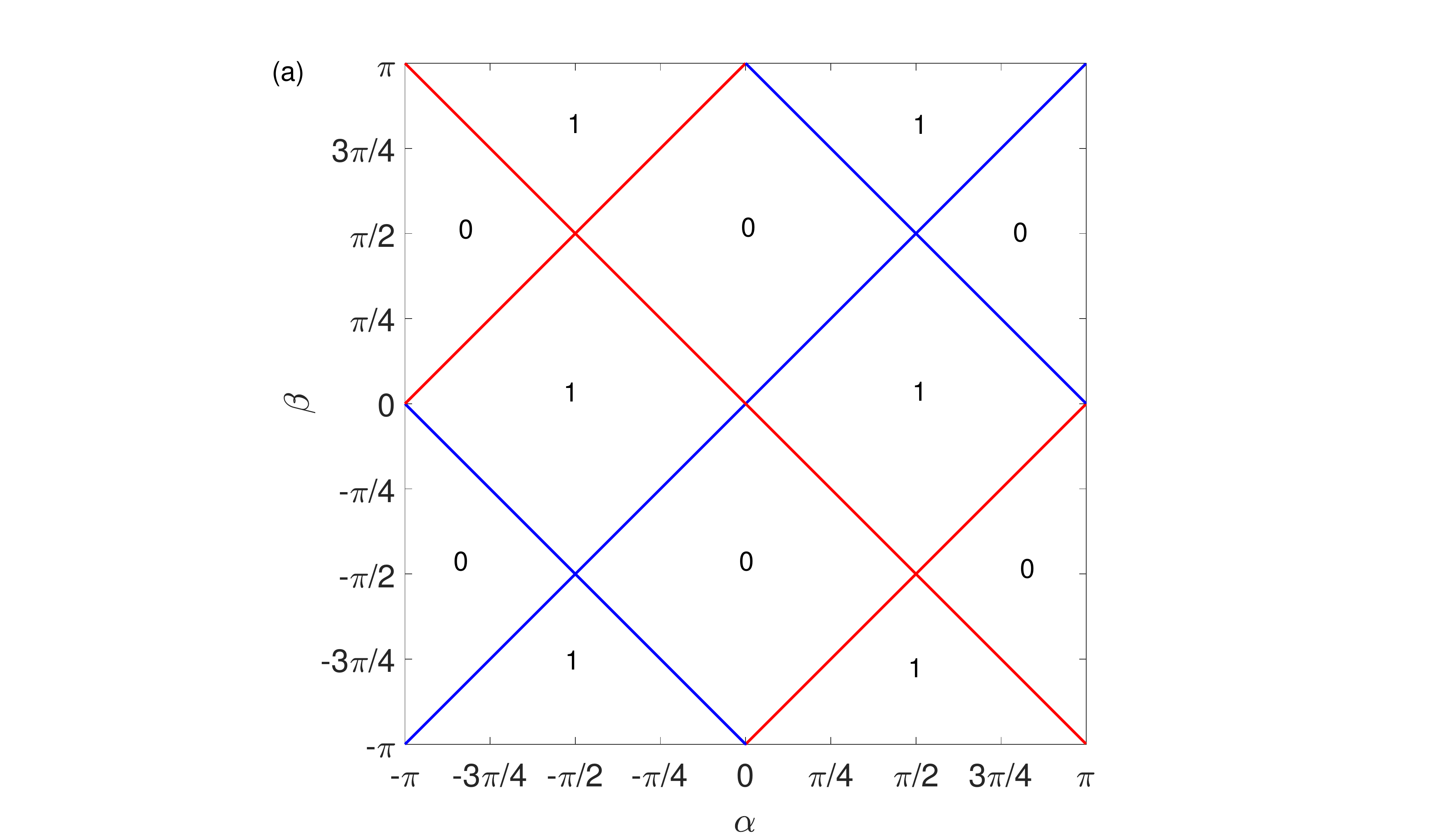}
\end{minipage}}
\quad
\subfigure{
\begin{minipage}[t]{0.4\linewidth}
\centering
\includegraphics[width=3in]{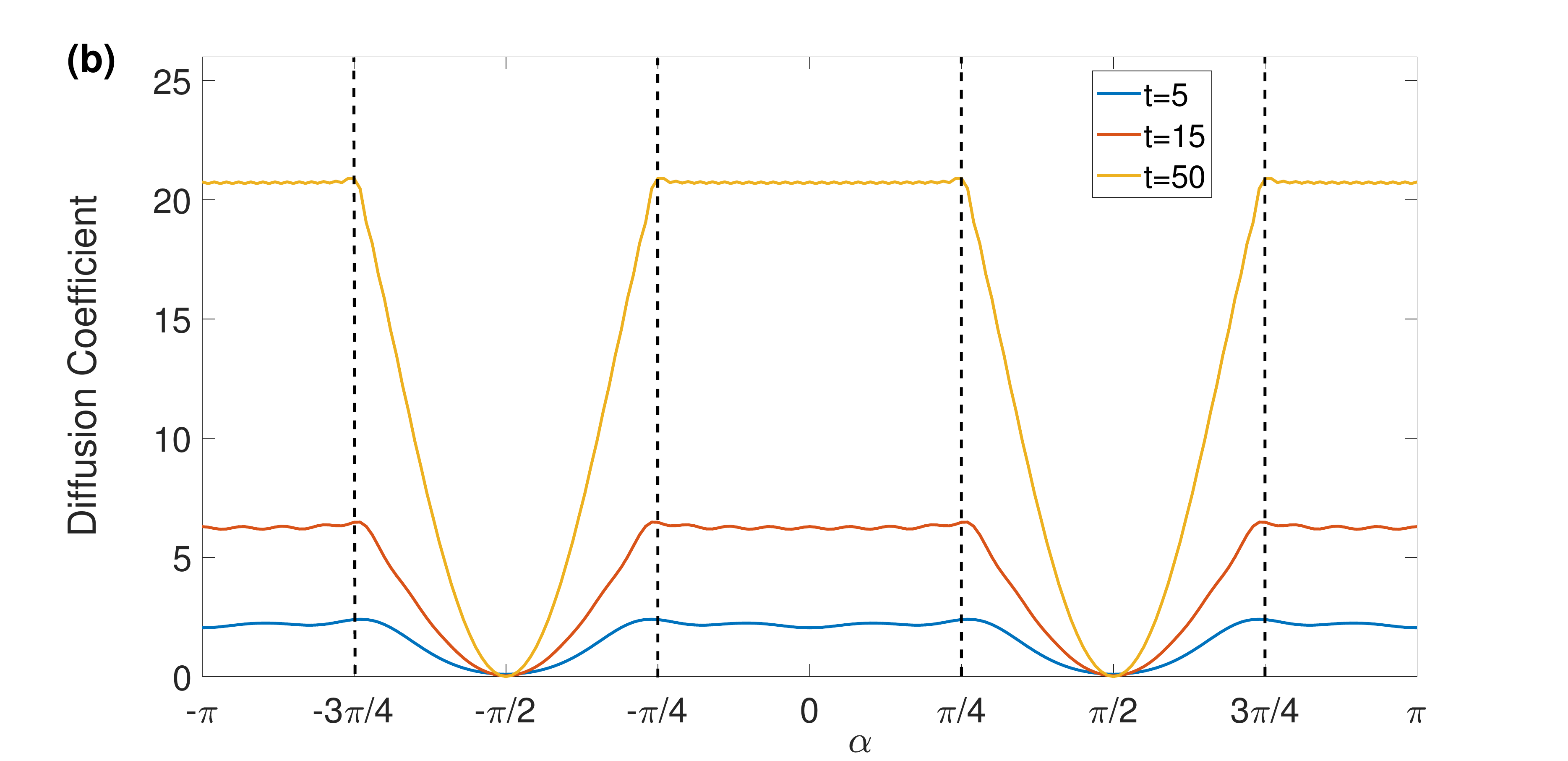}
\end{minipage}}
\quad
\subfigure{
\begin{minipage}[t]{0.4\linewidth}
\centering
\includegraphics[width=3in]{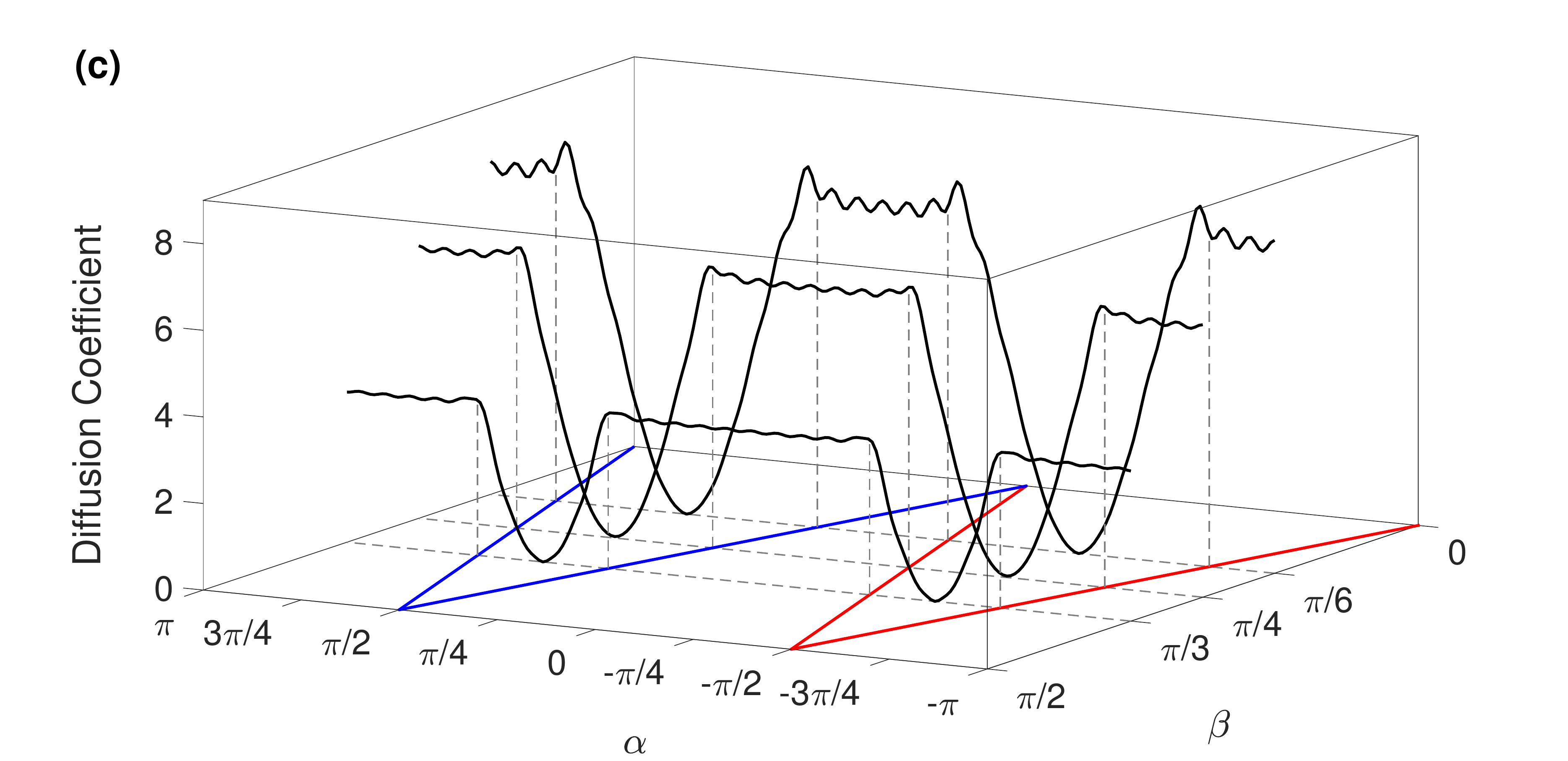}
\end{minipage}}
\quad
\subfigure{
\begin{minipage}[t]{0.4\linewidth}
\centering
\includegraphics[width=3in]{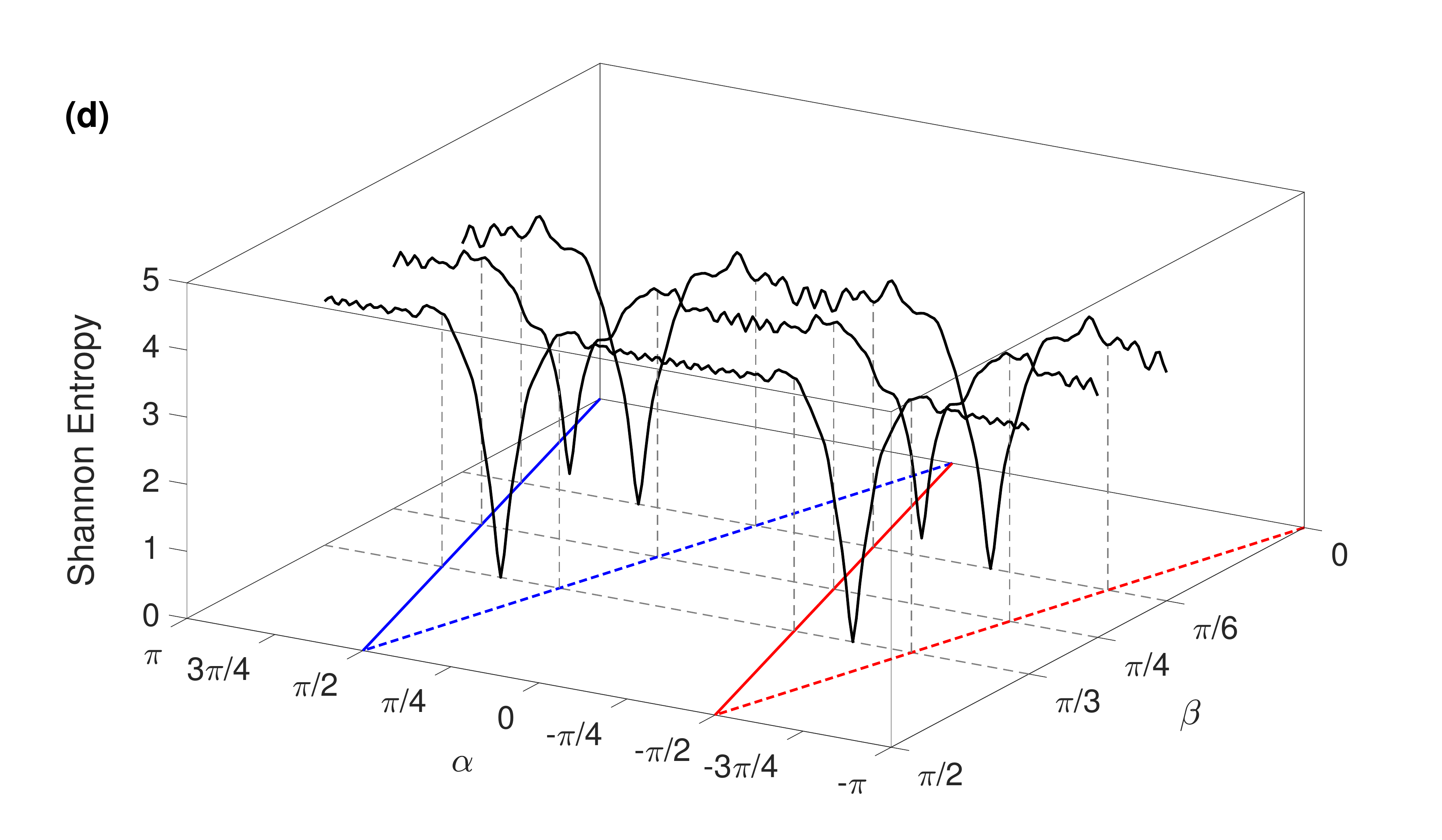}
\end{minipage}}
\centering
\caption{ (a) Phase diagram for discrete time quantum walk in unitary evolution, where winding
numbers are function of the spin-rotation angles $\alpha$ and $\beta$. Topologically distinct gapped
phases characterized by the winding number W are separated by phase-transition lines where the
quasienergy gap closes at either $E = 0$ (red) or $E = \pi$ (blue). (b) The diffusion coefficient of the
quantum walk in unitary evolution with different number of steps for the same $\beta = \pi/4$. Blue, red
and yellow lines correspond to $t = 5$, $t = 15$ and $t = 50$ steps, respectively. (c) The dependence
of diffusion coefficient on the spin-rotation angles $\alpha$ and $\beta$ with number of steps $t = 15$. (d)
The dependence of Shannon entropy on the spin-rotation angles $\alpha$ and $\beta$ with number of steps
$t = 15$.}
\label{Hermitian}
\end{figure*}

First, we study quantum walk with unitary evolution in Hermitian case. The quasi-energy of
the system and the Bloch vector are given by:
\begin{equation}
\cos \epsilon(k)=-\sin \alpha \sin \beta+\cos k \cos \alpha \cos \beta,
\end{equation}
and $\vec{n}(k)=n_{x}(k) \hat{i}+n_{y}(k) \hat{j}+n_{z}(k) \hat{k}$ with
\begin{equation}
\begin{array}{l}
n_{x}(k)=\frac{\sin k \sin \alpha \cos \beta}{\sin \epsilon(k)} \\
n_{y}(k)=\frac{\cos \alpha \sin \beta+\cos k \sin \alpha \cos \beta}{\sin \epsilon(k)} \\
n_{z}(k)=\frac{-\sin k \cos \alpha \cos \beta}{\sin \epsilon(k)}
\end{array}
\end{equation}
Thus, we get the winding number with $\Gamma=\left(\cos \alpha, 0, \sin \alpha\right)$ according to:
\begin{equation}
W=\frac{1}{2 \pi} \oint d k\left(\vec{n} \times \frac{\partial \vec{n}}{\partial k}\right) \cdot \Gamma
\end{equation}
The resulted phase diagram for the winding numbers is shown in Fig.\ref{Hermitian}(a). We observe two
topological phases corresponding to $W = 0$ and $W = 1$ which are separated by phase-transition
lines. Red and blue lines represent closing of the quasienergy gap at $\epsilon = 0$ and $\epsilon = \pi$, respectively.
We simulate the topology of the quantum walk on a one-dimensional lattice. Let the initial state
of the walker as $|\psi\rangle=|0\rangle \otimes|\uparrow\rangle$, i.e. the initial position of the walker is $x = 0$, initial coil state is $| \uparrow\rangle$.

Then, the calculation of the diffusion coefficient is carried out with different number of steps $t$.
The diffusion coefficient D is calculated according to \cite{romanelli2005decoherence,zhang2017topology}:
\begin{equation}
D=\frac{\sigma^{2}(t)}{2 t}=\frac{M_{2}(t)-M_{1}^{2}(t)}{2 t}
\end{equation}
where $M_{1}(t)=\sum_{n=-t}^{n=t} n P_{n}(t)$ and $M_{2}(t)=\sum_{n=-t}^{n=t} n^2 P_{n}(t)$ are the first and second moment, $P_{n}(t)$
is the probability that the position of the walker is n at time t. $\sigma^{2}$ is the variance of the quantum
walk. Fig.\ref{Hermitian}(b) shows the results of $t = 5$, $t = 15$ and $t = 50$ with the same $\beta = \pi/4$. For all
three chosen t, they show a very similar trend with $\alpha$, although the diffusion coefficient increases
with t. In the topological phase of $W = 0$, the diffusion coefficient is quite stable with small
fluctuations, while in the topological phase of $W = 1$, it shows a valley structure along with $\alpha$.
More importantly, in the close proximity of the topological phase transition, turning point appears
on the diffusion coefficient with the change of $\alpha$.
In Fig.\ref{Hermitian}(c), the dependence of diffusion coefficient on spin-rotation angle is given for
$\beta = \pi/6$, $\beta = \pi/4$ and $\beta = \pi/3$ with number of steps $t = 15$, respectively. On the bottom $\alpha\theta_2$
plane, phase-transition lines are shown as Fig.\ref{Hermitian}(a), together with vertical dashed line in order
to clearly distinguish those turning points in the vicinity to the topological phase transition. It is
also noticed that, diffusion coefficient decreases with the increase of $\beta$ for the same $\alpha$.

Subsequently, we obtain the corresponding Shannon entropy of the walker according to:
\begin{equation}
S=-\sum_{i=1}^{m} P_{n}\left(x_{i}\right) \log _{2} P_{n}\left(x_{i}\right)
\end{equation}
where $P_{n}\left(x_{i}\right)$ is the probability that a quantum walker is in position $x_{i}$ at time $n$. The results are
illustrated in Fig.\ref{Hermitian}(d). It is clear and interesting to find out, the Shannon entropy resembles
amazingly with the diffusion coefficient in terms of the dependence on spin-rotation angles $\alpha$ and
$\beta$, probably due to both of them are related to $P_{n}(t)$.

\subsection{PT-Symmetric Non-Hermitian Quantum Walk}

\begin{figure*}
\centering
\subfigure{
\begin{minipage}{12cm}
\centering
\includegraphics[width=0.9\textwidth]{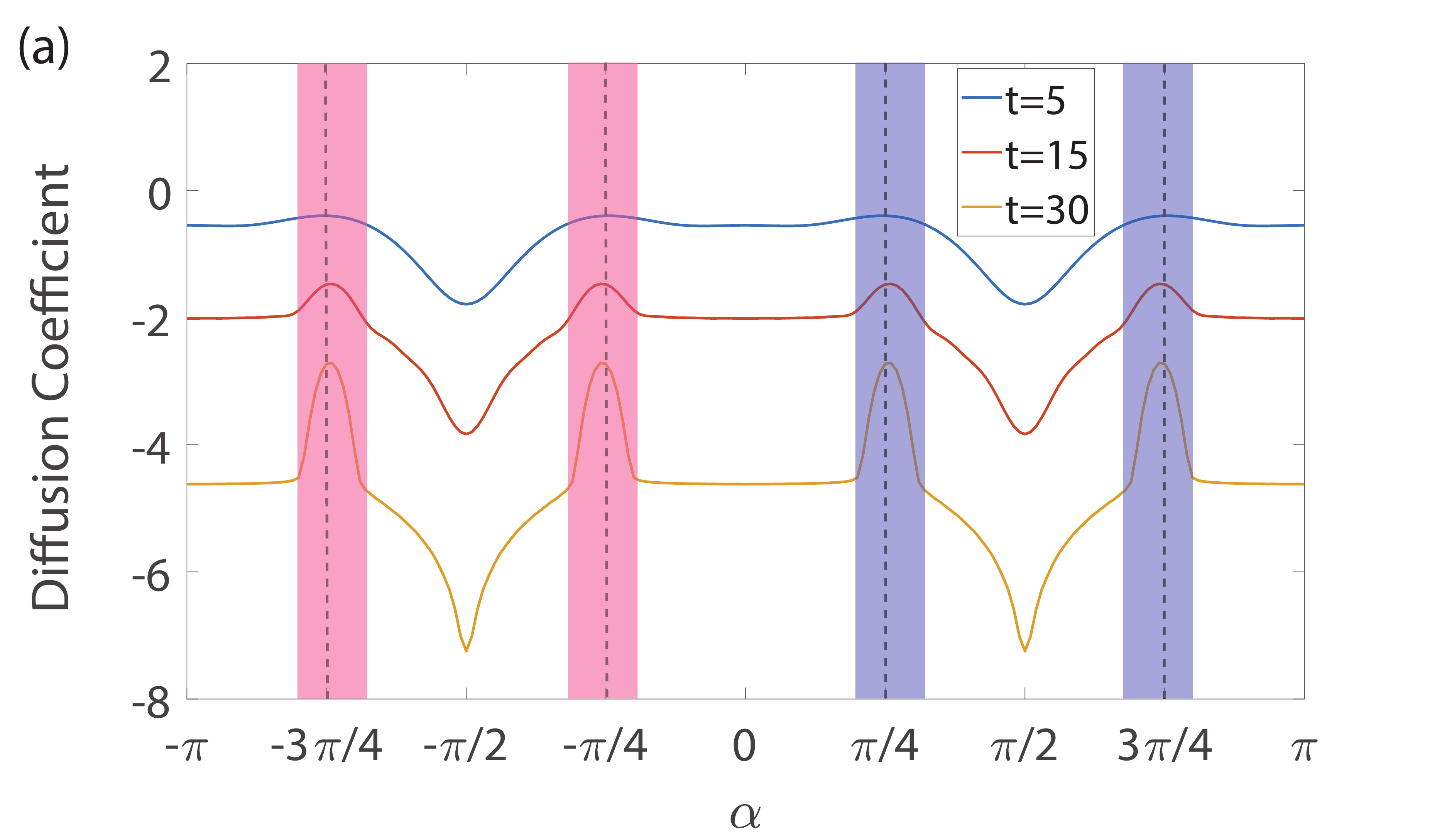} \\
\end{minipage}}

\subfigure{
\begin{minipage}{12cm}
\centering
\includegraphics[width=0.9\textwidth]{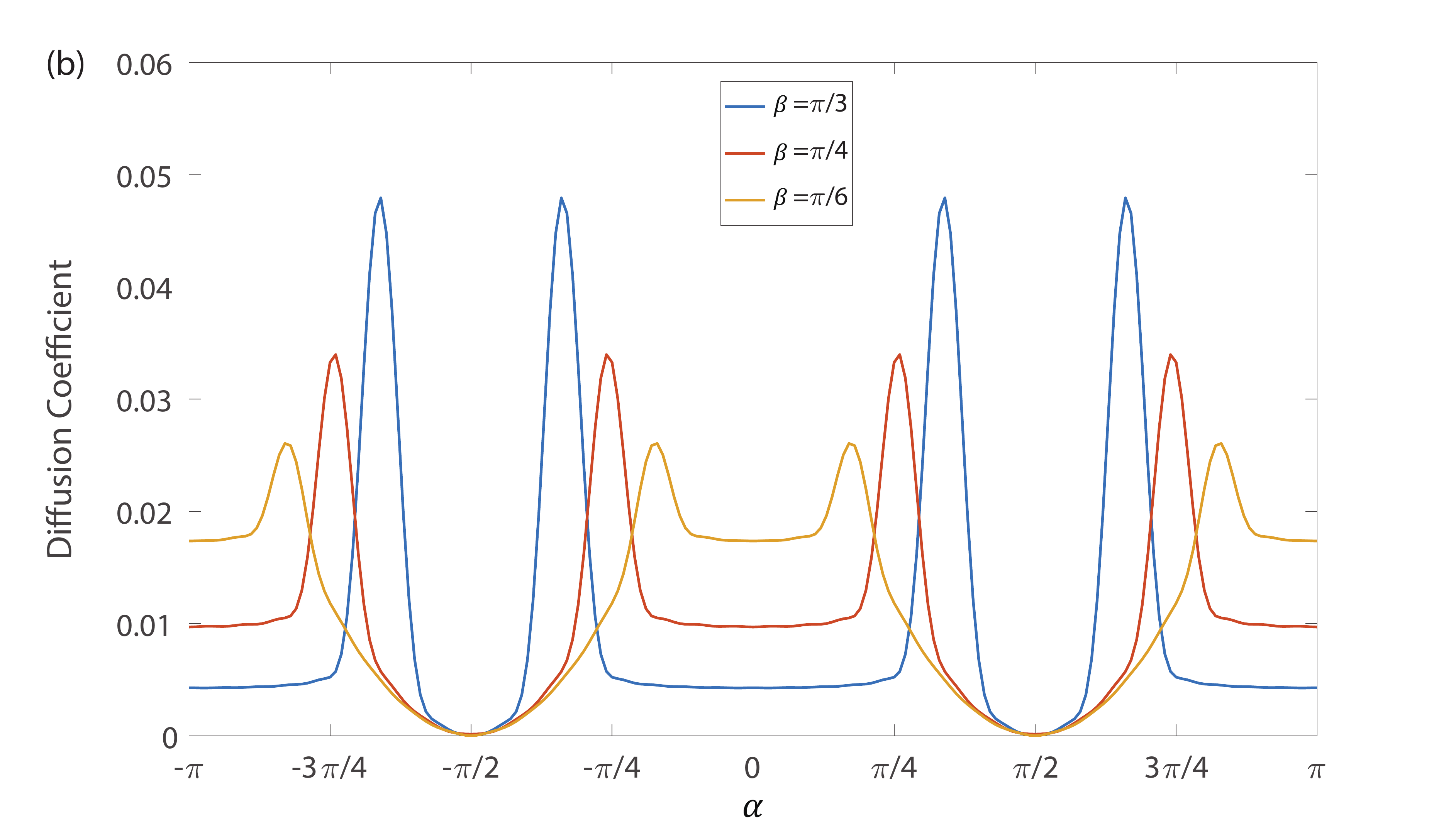} \\
\end{minipage}}

\subfigure{
\begin{minipage}{12cm}
\centering
\includegraphics[width=0.9\textwidth]{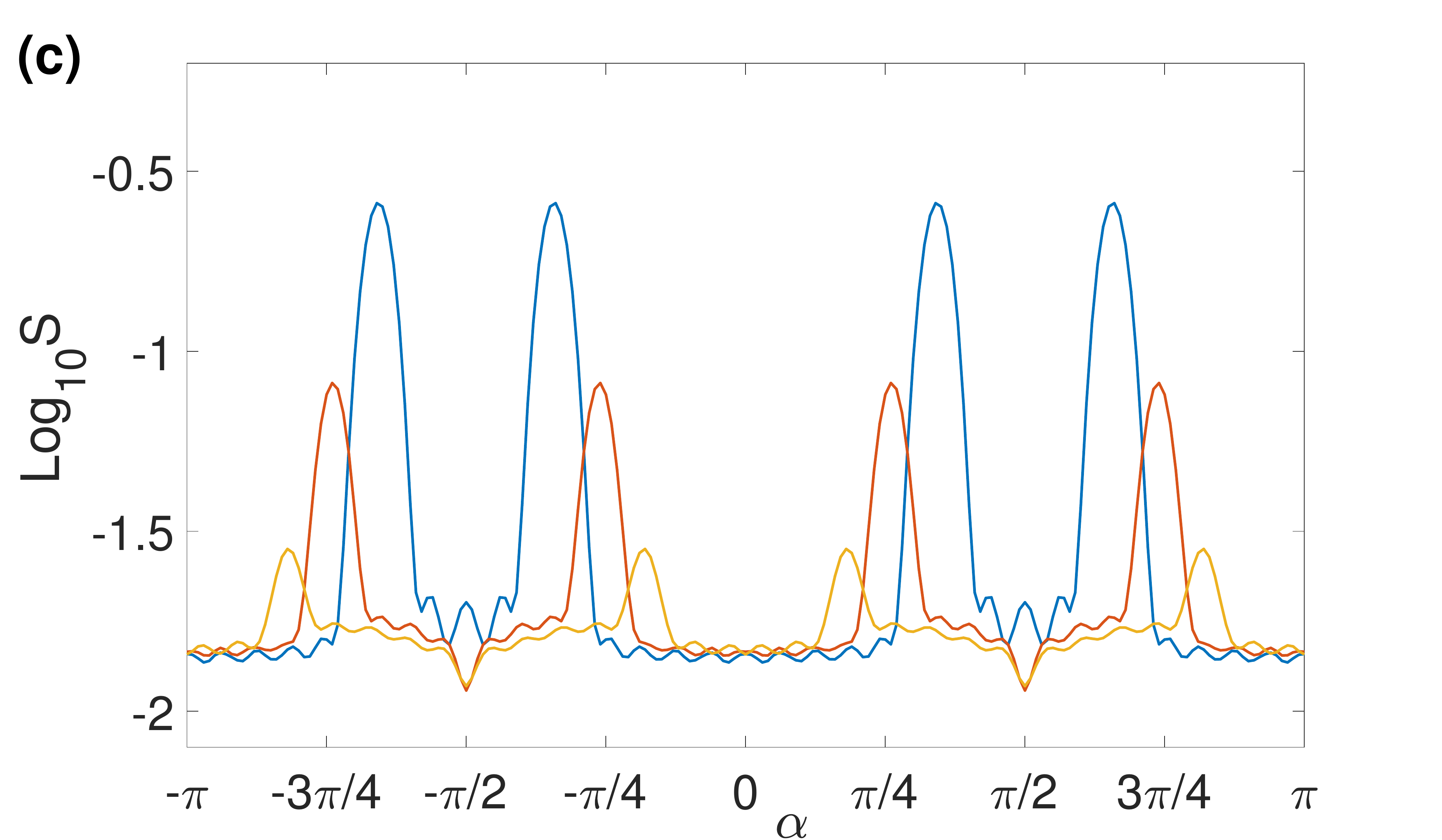} \\
\end{minipage}}
\caption{(a) The diffusion coefficient of the quantum walk in PT-symmetric non-unitary evolution with different number of steps for the same $\beta = \pi/4$. Blue, red and yellow lines correspond to $t = 5$, $t = 15$ and $t = 30$ steps, respectively. The blue and red shaded regions represent broken PT-symmetric phases \cite{xiao2017observation} with complex eigenenergies whose real parts are 0 and $\pi$, respectively. (b) The dependence of diffusion coefficient on the spin-rotation angle $\alpha$ with number of steps $t = 15$ for $\beta = \pi/6$, $\beta = \pi/4$ and $\beta = \pi/3$ respectively. (c) The dependence of Shannon entropy on the spin-rotation angle $\alpha$ with number of steps $t = 15$ for $\beta = \pi/6$, $\beta = \pi/4$ and $\beta = \pi/3$ respectively.}
\label{non-Hermitian}
\end{figure*}

In particular, we are more interested in the behavior of a non-Hermitian quantum walk, where the quantum coin (or position) carried by the walker interacts with an environment, resulting in a loss. Such system would allow us to explore topological properties of non-Hermitian systems and to investigate Floquet topological phases \cite{mochizuki2016explicit, xiao2017observation}. Moreover, it is shown that the quantum walk properties are highly sensitive to decoherent events. As the quantum coil (or position) is coupled to an environment, decoherence would lead a transition from quantum walk to classical random walk for large number of steps, and a relation between the decoherence and the diffusion properties of the system can be established \cite{brun2003quantum,lopez2003phase,kendon2003decoherence,romanelli2005decoherence}. In addition, a weak decoherence could possibly enhance the properties of the quantum walk and is thus beneficial to the development of quantum algorithms \cite{kendon2003decoherence}.

Here we consider the PT-symmetric non-Hermitian scenario by calculating $\left(l_{1}, l_{2}\right)=(1,0.8)$ with the
same calculation and analysis. We calculate diffusion coefficients for different number of steps for
$t = 5$, $t = 15$ and $t = 30$ three different number of steps with the same $\beta = \pi/4$, the results are
shown in Fig.\ref{non-Hermitian}(a). Based on the phase diagram of the PT-symmetric non-Hermitian quantum
walk \cite{xiao2017observation}, the diffusion coefficient are found to show similar features to Fig.\ref{Hermitian}(b), which is
exhibiting abrupt change at the boundary of the topological phase transition. As $\alpha$ increases
from $-\pi$ to 0, the system starts from PT-symmetric phase (corresponding to a topologically
non-trivial state with topological number $(\nu_0, \nu_\pi) = (1, 1)$), and transits broken PT-symmetric
phase which is represented by red region in the vicinity of $\alpha = -3\pi/4$, with which the real
part of the eigenenergy is $\pi$. Then the system enters another PT-symmetric phase (topologically
non-trivial state with topological number $(\nu_0, \nu_\pi) = (1, -1)$), and transits previous broken PTsymmetric
phase again, finally reaches a PT-symmetric phase (topologically non-trivial state with
topological number $(\nu_0, \nu_\pi) = (1, 1)$). The diffusion coefficient stays stable until it reaches the
broken PT-symmetric phase. When it first crosses broken PT-symmetric phase, the diffusion
coefficient gradually increases before $\alpha = -3\pi/4$ and then decreases, exhibiting a peak-valley
profile. When it crosses broken PT-symmetric phase again, the diffusion coefficient increases
before $\alpha = -\pi/4$ and then decreases, exhibiting a valley-peak profile. The varies of the diffusion
coefficient from $\alpha = 0$ to $\pi$ followes the same way as from $\alpha = -\pi$ to 0.

Another feature of the diffusion coefficient is that its amplitude is much smaller than that in the
Hermitian case, and it decreases with the increase of the number of the step t due to the existing
of the loss operator. Fig.\ref{non-Hermitian}(b) illustrates the dependence of diffusion coefficient on spin-rotation
angle $\alpha$ for $\beta = \pi/6$, $\beta = \pi/4$ and $\beta = \pi/3$ with number of steps $t = 15$, respectively. We choose
$t = 15$ because of several reasons. First, we find the effects by varying t is much larger than by
varying $\alpha$. Second, larger $t$ would lead to too small the value of diffusion coefficient to observe
any effect due to excessive loss. Third, the effect of coherent diffusion would be invisable if the
number of steps t is very small. Therefore, in order to better observe the effect of PT-symmetric
quantum walk with the topological phase in the non-Hermitian scenario, we take $t = 15$ for the
following calculations for both diffusion coefficient and Shannon entropy.
From Fig.\ref{non-Hermitian}(a-b), we notice that, the peak profile of the diffusion coefficient
has direct correspondence to the broken PT-symmetric phase, i.e. the topological edge state, which suggests a possible intrinsic relation between them. In Fig.\ref{non-Hermitian}(c), Shannon entropy as the
function of spin-rotation angle $\alpha$ for $\beta = \pi/6$, $\beta = \pi/4$ and $\beta = \pi/3$ with number of steps $t = 15$
are presented, respectively. It is obvious to find, all of them exhibit an interesting resemblance to
that for diffusion coefficient.

\section{Conclusion}

We have proposed using the diffusion coefficient as a probe for studying the dynamics of topological
edge states. We have conducted simulations of both Hermitian quantum walk and PTsymmetric
non-Hermitian quantum walk, and have investigated the topological phases driven by
such quantum walks. The properties of diffusion coefficient as a function of two parameters $\alpha$
and $\beta$ in evolution operator of the quantum walk are studied. It has been found that that diffusion
coefficient exhibits qualitatively similar characteristics as the topological phase transition
appears in Hermitian and PT-symmetric non-Hermitian quantum walks. For a Hermitian system,
a turning point and abrupt change appears in the diffusion coefficient when the system is
at the boundary of the topological phase transition, while it remains stable in trivial topological
state. For a PT-symmetric non-Hermitian system, however, the diffusion coefficient displays unique peak
structure in the topological phase transition region. In addition, Shannon entropy is found to
illustrate very corresponding behavior as the diffusion coefficient. Such interesting signatures might give
us a unique probe for the topological edge states, and suggest exciting possibilities for studying
the topological boundary effects and edge states in PT-symmetric non-Hermitian quantum walks
by exploiting diffusion coefficient and Shannon entropy.

%\textcolor{red}{Moreover, we envision one experiment that could be done in the near future, which drives single trapped $^{171}Yb^+$ ion with a pair of Raman beams to control the ion¡¯s momentum state and uses the ion¡¯s momentum state and internal hyperfine energy levels in ground electronic state as the position and coin state, respectively. A dissipation laser which drives an electronic transition can induce the particle loss in the hyperfine ion qubit and create a Parity-Time symmetric open system. With this configuration, the diffusion coefficient and the Shannon entropy simulated in this letter can be experimentally investigated. }

\section*{Acknowledgements}

This work is supported by the Key-Area Research and Development Program
of Guang Dong Province under Grant No.2019B030330001, the National Natural Science Foundation of China under Grant
No.11774436 and No.11974434. Le Luo received supports from Guangdong Province Youth Talent Program under Grant No.2017GC010656, Sun Yat-sen University Core Technology Development Fund. Yang Liu acknowledges the support from Natural Science Foundation of Guangdong Province under Grant 2020A1515011159 and Fundamental Research Funds for the Central Universities
of Education of China under Grant No. 191gpy276.

\bibliography{sample2}

%merlin.mbs apsrev4-1.bst 2010-07-25 4.21a (PWD, AO, DPC) hacked
%Control: key (0)
%Control: author (8) initials jnrlst
%Control: editor formatted (1) identically to author
%Control: production of article title (-1) disabled
%Control: page (0) single
%Control: year (1) truncated
%Control: production of eprint (0) enabled
\begin{thebibliography}{41}%
\makeatletter
\providecommand \@ifxundefined [1]{%
 \@ifx{#1\undefined}
}%
\providecommand \@ifnum [1]{%
 \ifnum #1\expandafter \@firstoftwo
 \else \expandafter \@secondoftwo
 \fi
}%
\providecommand \@ifx [1]{%
 \ifx #1\expandafter \@firstoftwo
 \else \expandafter \@secondoftwo
 \fi
}%
\providecommand \natexlab [1]{#1}%
\providecommand \enquote  [1]{``#1''}%
\providecommand \bibnamefont  [1]{#1}%
\providecommand \bibfnamefont [1]{#1}%
\providecommand \citenamefont [1]{#1}%
\providecommand \href@noop [0]{\@secondoftwo}%
\providecommand \href [0]{\begingroup \@sanitize@url \@href}%
\providecommand \@href[1]{\@@startlink{#1}\@@href}%
\providecommand \@@href[1]{\endgroup#1\@@endlink}%
\providecommand \@sanitize@url [0]{\catcode `\\12\catcode `\$12\catcode
  `\&12\catcode `\#12\catcode `\^12\catcode `\_12\catcode `\%12\relax}%
\providecommand \@@startlink[1]{}%
\providecommand \@@endlink[0]{}%
\providecommand \url  [0]{\begingroup\@sanitize@url \@url }%
\providecommand \@url [1]{\endgroup\@href {#1}{\urlprefix }}%
\providecommand \urlprefix  [0]{URL }%
\providecommand \Eprint [0]{\href }%
\providecommand \doibase [0]{http://dx.doi.org/}%
\providecommand \selectlanguage [0]{\@gobble}%
\providecommand \bibinfo  [0]{\@secondoftwo}%
\providecommand \bibfield  [0]{\@secondoftwo}%
\providecommand \translation [1]{[#1]}%
\providecommand \BibitemOpen [0]{}%
\providecommand \bibitemStop [0]{}%
\providecommand \bibitemNoStop [0]{.\EOS\space}%
\providecommand \EOS [0]{\spacefactor3000\relax}%
\providecommand \BibitemShut  [1]{\csname bibitem#1\endcsname}%
\let\auto@bib@innerbib\@empty
%</preamble>
\bibitem [{\citenamefont {Childs}(2009)}]{childs2009universal}%
  \BibitemOpen
  \bibfield  {author} {\bibinfo {author} {\bibfnamefont {A.~M.}\ \bibnamefont
  {Childs}},\ }\href@noop {} {\bibfield  {journal} {\bibinfo  {journal}
  {Physical review letters}\ }\textbf {\bibinfo {volume} {102}},\ \bibinfo
  {pages} {180501} (\bibinfo {year} {2009})}\BibitemShut {NoStop}%
\bibitem [{\citenamefont {Karski}\ \emph {et~al.}(2009)\citenamefont {Karski},
  \citenamefont {F{\"o}rster}, \citenamefont {Choi}, \citenamefont {Steffen},
  \citenamefont {Alt}, \citenamefont {Meschede},\ and\ \citenamefont
  {Widera}}]{karski2009quantum}%
  \BibitemOpen
  \bibfield  {author} {\bibinfo {author} {\bibfnamefont {M.}~\bibnamefont
  {Karski}}, \bibinfo {author} {\bibfnamefont {L.}~\bibnamefont {F{\"o}rster}},
  \bibinfo {author} {\bibfnamefont {J.-M.}\ \bibnamefont {Choi}}, \bibinfo
  {author} {\bibfnamefont {A.}~\bibnamefont {Steffen}}, \bibinfo {author}
  {\bibfnamefont {W.}~\bibnamefont {Alt}}, \bibinfo {author} {\bibfnamefont
  {D.}~\bibnamefont {Meschede}}, \ and\ \bibinfo {author} {\bibfnamefont
  {A.}~\bibnamefont {Widera}},\ }\href@noop {} {\bibfield  {journal} {\bibinfo
  {journal} {Science}\ }\textbf {\bibinfo {volume} {325}},\ \bibinfo {pages}
  {174} (\bibinfo {year} {2009})}\BibitemShut {NoStop}%
\bibitem [{\citenamefont {Schmitz}\ \emph {et~al.}(2009)\citenamefont
  {Schmitz}, \citenamefont {Matjeschk}, \citenamefont {Schneider},
  \citenamefont {Glueckert}, \citenamefont {Enderlein}, \citenamefont {Huber},\
  and\ \citenamefont {Schaetz}}]{schmitz2009quantum}%
  \BibitemOpen
  \bibfield  {author} {\bibinfo {author} {\bibfnamefont {H.}~\bibnamefont
  {Schmitz}}, \bibinfo {author} {\bibfnamefont {R.}~\bibnamefont {Matjeschk}},
  \bibinfo {author} {\bibfnamefont {C.}~\bibnamefont {Schneider}}, \bibinfo
  {author} {\bibfnamefont {J.}~\bibnamefont {Glueckert}}, \bibinfo {author}
  {\bibfnamefont {M.}~\bibnamefont {Enderlein}}, \bibinfo {author}
  {\bibfnamefont {T.}~\bibnamefont {Huber}}, \ and\ \bibinfo {author}
  {\bibfnamefont {T.}~\bibnamefont {Schaetz}},\ }\href@noop {} {\bibfield
  {journal} {\bibinfo  {journal} {Physical review letters}\ }\textbf {\bibinfo
  {volume} {103}},\ \bibinfo {pages} {090504} (\bibinfo {year}
  {2009})}\BibitemShut {NoStop}%
\bibitem [{\citenamefont {Lovett}\ \emph {et~al.}(2010)\citenamefont {Lovett},
  \citenamefont {Cooper}, \citenamefont {Everitt}, \citenamefont {Trevers},\
  and\ \citenamefont {Kendon}}]{lovett2010universal}%
  \BibitemOpen
  \bibfield  {author} {\bibinfo {author} {\bibfnamefont {N.~B.}\ \bibnamefont
  {Lovett}}, \bibinfo {author} {\bibfnamefont {S.}~\bibnamefont {Cooper}},
  \bibinfo {author} {\bibfnamefont {M.}~\bibnamefont {Everitt}}, \bibinfo
  {author} {\bibfnamefont {M.}~\bibnamefont {Trevers}}, \ and\ \bibinfo
  {author} {\bibfnamefont {V.}~\bibnamefont {Kendon}},\ }\href@noop {}
  {\bibfield  {journal} {\bibinfo  {journal} {Physical Review A}\ }\textbf
  {\bibinfo {volume} {81}},\ \bibinfo {pages} {042330} (\bibinfo {year}
  {2010})}\BibitemShut {NoStop}%
\bibitem [{\citenamefont {Z{\"a}hringer}\ \emph {et~al.}(2010)\citenamefont
  {Z{\"a}hringer}, \citenamefont {Kirchmair}, \citenamefont {Gerritsma},
  \citenamefont {Solano}, \citenamefont {Blatt},\ and\ \citenamefont
  {Roos}}]{zahringer2010realization}%
  \BibitemOpen
  \bibfield  {author} {\bibinfo {author} {\bibfnamefont {F.}~\bibnamefont
  {Z{\"a}hringer}}, \bibinfo {author} {\bibfnamefont {G.}~\bibnamefont
  {Kirchmair}}, \bibinfo {author} {\bibfnamefont {R.}~\bibnamefont
  {Gerritsma}}, \bibinfo {author} {\bibfnamefont {E.}~\bibnamefont {Solano}},
  \bibinfo {author} {\bibfnamefont {R.}~\bibnamefont {Blatt}}, \ and\ \bibinfo
  {author} {\bibfnamefont {C.}~\bibnamefont {Roos}},\ }\href@noop {} {\bibfield
   {journal} {\bibinfo  {journal} {Physical review letters}\ }\textbf {\bibinfo
  {volume} {104}},\ \bibinfo {pages} {100503} (\bibinfo {year}
  {2010})}\BibitemShut {NoStop}%
\bibitem [{\citenamefont {Childs}\ \emph {et~al.}(2013)\citenamefont {Childs},
  \citenamefont {Gosset},\ and\ \citenamefont {Webb}}]{childs2013universal}%
  \BibitemOpen
  \bibfield  {author} {\bibinfo {author} {\bibfnamefont {A.~M.}\ \bibnamefont
  {Childs}}, \bibinfo {author} {\bibfnamefont {D.}~\bibnamefont {Gosset}}, \
  and\ \bibinfo {author} {\bibfnamefont {Z.}~\bibnamefont {Webb}},\ }\href@noop
  {} {\bibfield  {journal} {\bibinfo  {journal} {Science}\ }\textbf {\bibinfo
  {volume} {339}},\ \bibinfo {pages} {791} (\bibinfo {year}
  {2013})}\BibitemShut {NoStop}%
\bibitem [{\citenamefont {Perets}\ \emph {et~al.}(2008)\citenamefont {Perets},
  \citenamefont {Lahini}, \citenamefont {Pozzi}, \citenamefont {Sorel},
  \citenamefont {Morandotti},\ and\ \citenamefont
  {Silberberg}}]{perets2008realization}%
  \BibitemOpen
  \bibfield  {author} {\bibinfo {author} {\bibfnamefont {H.~B.}\ \bibnamefont
  {Perets}}, \bibinfo {author} {\bibfnamefont {Y.}~\bibnamefont {Lahini}},
  \bibinfo {author} {\bibfnamefont {F.}~\bibnamefont {Pozzi}}, \bibinfo
  {author} {\bibfnamefont {M.}~\bibnamefont {Sorel}}, \bibinfo {author}
  {\bibfnamefont {R.}~\bibnamefont {Morandotti}}, \ and\ \bibinfo {author}
  {\bibfnamefont {Y.}~\bibnamefont {Silberberg}},\ }\href@noop {} {\bibfield
  {journal} {\bibinfo  {journal} {Physical review letters}\ }\textbf {\bibinfo
  {volume} {100}},\ \bibinfo {pages} {170506} (\bibinfo {year}
  {2008})}\BibitemShut {NoStop}%
\bibitem [{\citenamefont {Broome}\ \emph {et~al.}(2010)\citenamefont {Broome},
  \citenamefont {Fedrizzi}, \citenamefont {Lanyon}, \citenamefont {Kassal},
  \citenamefont {Aspuru-Guzik},\ and\ \citenamefont
  {White}}]{broome2010discrete}%
  \BibitemOpen
  \bibfield  {author} {\bibinfo {author} {\bibfnamefont {M.~A.}\ \bibnamefont
  {Broome}}, \bibinfo {author} {\bibfnamefont {A.}~\bibnamefont {Fedrizzi}},
  \bibinfo {author} {\bibfnamefont {B.~P.}\ \bibnamefont {Lanyon}}, \bibinfo
  {author} {\bibfnamefont {I.}~\bibnamefont {Kassal}}, \bibinfo {author}
  {\bibfnamefont {A.}~\bibnamefont {Aspuru-Guzik}}, \ and\ \bibinfo {author}
  {\bibfnamefont {A.~G.}\ \bibnamefont {White}},\ }\href@noop {} {\bibfield
  {journal} {\bibinfo  {journal} {Physical review letters}\ }\textbf {\bibinfo
  {volume} {104}},\ \bibinfo {pages} {153602} (\bibinfo {year}
  {2010})}\BibitemShut {NoStop}%
\bibitem [{\citenamefont {Thouless}\ \emph {et~al.}(1982)\citenamefont
  {Thouless}, \citenamefont {Kohmoto}, \citenamefont {Nightingale},\ and\
  \citenamefont {den Nijs}}]{thouless1982quantized}%
  \BibitemOpen
  \bibfield  {author} {\bibinfo {author} {\bibfnamefont {D.~J.}\ \bibnamefont
  {Thouless}}, \bibinfo {author} {\bibfnamefont {M.}~\bibnamefont {Kohmoto}},
  \bibinfo {author} {\bibfnamefont {M.~P.}\ \bibnamefont {Nightingale}}, \ and\
  \bibinfo {author} {\bibfnamefont {M.}~\bibnamefont {den Nijs}},\ }\href@noop
  {} {\bibfield  {journal} {\bibinfo  {journal} {Physical review letters}\
  }\textbf {\bibinfo {volume} {49}},\ \bibinfo {pages} {405} (\bibinfo {year}
  {1982})}\BibitemShut {NoStop}%
\bibitem [{\citenamefont {Kitagawa}\ \emph {et~al.}(2010)\citenamefont
  {Kitagawa}, \citenamefont {Rudner}, \citenamefont {Berg},\ and\ \citenamefont
  {Demler}}]{kitagawa2010exploring}%
  \BibitemOpen
  \bibfield  {author} {\bibinfo {author} {\bibfnamefont {T.}~\bibnamefont
  {Kitagawa}}, \bibinfo {author} {\bibfnamefont {M.~S.}\ \bibnamefont
  {Rudner}}, \bibinfo {author} {\bibfnamefont {E.}~\bibnamefont {Berg}}, \ and\
  \bibinfo {author} {\bibfnamefont {E.}~\bibnamefont {Demler}},\ }\href@noop {}
  {\bibfield  {journal} {\bibinfo  {journal} {Physical Review A}\ }\textbf
  {\bibinfo {volume} {82}},\ \bibinfo {pages} {033429} (\bibinfo {year}
  {2010})}\BibitemShut {NoStop}%
\bibitem [{\citenamefont {Kitagawa}\ \emph {et~al.}(2012)\citenamefont
  {Kitagawa}, \citenamefont {Broome}, \citenamefont {Fedrizzi}, \citenamefont
  {Rudner}, \citenamefont {Berg}, \citenamefont {Kassal}, \citenamefont
  {Aspuru-Guzik}, \citenamefont {Demler},\ and\ \citenamefont
  {White}}]{kitagawa2012observation}%
  \BibitemOpen
  \bibfield  {author} {\bibinfo {author} {\bibfnamefont {T.}~\bibnamefont
  {Kitagawa}}, \bibinfo {author} {\bibfnamefont {M.~A.}\ \bibnamefont
  {Broome}}, \bibinfo {author} {\bibfnamefont {A.}~\bibnamefont {Fedrizzi}},
  \bibinfo {author} {\bibfnamefont {M.~S.}\ \bibnamefont {Rudner}}, \bibinfo
  {author} {\bibfnamefont {E.}~\bibnamefont {Berg}}, \bibinfo {author}
  {\bibfnamefont {I.}~\bibnamefont {Kassal}}, \bibinfo {author} {\bibfnamefont
  {A.}~\bibnamefont {Aspuru-Guzik}}, \bibinfo {author} {\bibfnamefont
  {E.}~\bibnamefont {Demler}}, \ and\ \bibinfo {author} {\bibfnamefont {A.~G.}\
  \bibnamefont {White}},\ }\href@noop {} {\bibfield  {journal} {\bibinfo
  {journal} {Nature communications}\ }\textbf {\bibinfo {volume} {3}},\
  \bibinfo {pages} {1} (\bibinfo {year} {2012})}\BibitemShut {NoStop}%
\bibitem [{\citenamefont {Cardano}\ \emph {et~al.}(2017)\citenamefont
  {Cardano}, \citenamefont {D’Errico}, \citenamefont {Dauphin}, \citenamefont
  {Maffei}, \citenamefont {Piccirillo}, \citenamefont {de~Lisio}, \citenamefont
  {De~Filippis}, \citenamefont {Cataudella}, \citenamefont {Santamato},
  \citenamefont {Marrucci} \emph {et~al.}}]{cardano2017detection}%
  \BibitemOpen
  \bibfield  {author} {\bibinfo {author} {\bibfnamefont {F.}~\bibnamefont
  {Cardano}}, \bibinfo {author} {\bibfnamefont {A.}~\bibnamefont {D’Errico}},
  \bibinfo {author} {\bibfnamefont {A.}~\bibnamefont {Dauphin}}, \bibinfo
  {author} {\bibfnamefont {M.}~\bibnamefont {Maffei}}, \bibinfo {author}
  {\bibfnamefont {B.}~\bibnamefont {Piccirillo}}, \bibinfo {author}
  {\bibfnamefont {C.}~\bibnamefont {de~Lisio}}, \bibinfo {author}
  {\bibfnamefont {G.}~\bibnamefont {De~Filippis}}, \bibinfo {author}
  {\bibfnamefont {V.}~\bibnamefont {Cataudella}}, \bibinfo {author}
  {\bibfnamefont {E.}~\bibnamefont {Santamato}}, \bibinfo {author}
  {\bibfnamefont {L.}~\bibnamefont {Marrucci}},  \emph {et~al.},\ }\href@noop
  {} {\bibfield  {journal} {\bibinfo  {journal} {Nature communications}\
  }\textbf {\bibinfo {volume} {8}},\ \bibinfo {pages} {1} (\bibinfo {year}
  {2017})}\BibitemShut {NoStop}%
\bibitem [{\citenamefont {Attal}\ \emph
  {et~al.}(2012{\natexlab{a}})\citenamefont {Attal}, \citenamefont
  {Petruccione},\ and\ \citenamefont {Sinayskiy}}]{attal2012open}%
  \BibitemOpen
  \bibfield  {author} {\bibinfo {author} {\bibfnamefont {S.}~\bibnamefont
  {Attal}}, \bibinfo {author} {\bibfnamefont {F.}~\bibnamefont {Petruccione}},
  \ and\ \bibinfo {author} {\bibfnamefont {I.}~\bibnamefont {Sinayskiy}},\
  }\href@noop {} {\bibfield  {journal} {\bibinfo  {journal} {Physics Letters
  A}\ }\textbf {\bibinfo {volume} {376}},\ \bibinfo {pages} {1545} (\bibinfo
  {year} {2012}{\natexlab{a}})}\BibitemShut {NoStop}%
\bibitem [{\citenamefont {Attal}\ \emph
  {et~al.}(2012{\natexlab{b}})\citenamefont {Attal}, \citenamefont
  {Petruccione}, \citenamefont {Sabot},\ and\ \citenamefont
  {Sinayskiy}}]{attal2012open2}%
  \BibitemOpen
  \bibfield  {author} {\bibinfo {author} {\bibfnamefont {S.}~\bibnamefont
  {Attal}}, \bibinfo {author} {\bibfnamefont {F.}~\bibnamefont {Petruccione}},
  \bibinfo {author} {\bibfnamefont {C.}~\bibnamefont {Sabot}}, \ and\ \bibinfo
  {author} {\bibfnamefont {I.}~\bibnamefont {Sinayskiy}},\ }\href@noop {}
  {\bibfield  {journal} {\bibinfo  {journal} {Journal of Statistical Physics}\
  }\textbf {\bibinfo {volume} {147}},\ \bibinfo {pages} {832} (\bibinfo {year}
  {2012}{\natexlab{b}})}\BibitemShut {NoStop}%
\bibitem [{\citenamefont {Sinayskiy}\ and\ \citenamefont
  {Petruccione}(2013)}]{sinayskiy2013open}%
  \BibitemOpen
  \bibfield  {author} {\bibinfo {author} {\bibfnamefont {I.}~\bibnamefont
  {Sinayskiy}}\ and\ \bibinfo {author} {\bibfnamefont {F.}~\bibnamefont
  {Petruccione}},\ }in\ \href@noop {} {\emph {\bibinfo {booktitle} {Journal of
  Physics: Conference Series}}},\ Vol.\ \bibinfo {volume} {442}\ (\bibinfo
  {organization} {IOP Publishing},\ \bibinfo {year} {2013})\ p.\ \bibinfo
  {pages} {012003}\BibitemShut {NoStop}%
\bibitem [{\citenamefont {Bauer}\ \emph {et~al.}(2013)\citenamefont {Bauer},
  \citenamefont {Bernard},\ and\ \citenamefont {Tilloy}}]{bauer2013open}%
  \BibitemOpen
  \bibfield  {author} {\bibinfo {author} {\bibfnamefont {M.}~\bibnamefont
  {Bauer}}, \bibinfo {author} {\bibfnamefont {D.}~\bibnamefont {Bernard}}, \
  and\ \bibinfo {author} {\bibfnamefont {A.}~\bibnamefont {Tilloy}},\
  }\href@noop {} {\bibfield  {journal} {\bibinfo  {journal} {Physical Review
  A}\ }\textbf {\bibinfo {volume} {88}},\ \bibinfo {pages} {062340} (\bibinfo
  {year} {2013})}\BibitemShut {NoStop}%
\bibitem [{\citenamefont {Sinayskiy}\ and\ \citenamefont
  {Petruccione}(2012)}]{sinayskiy2012efficiency}%
  \BibitemOpen
  \bibfield  {author} {\bibinfo {author} {\bibfnamefont {I.}~\bibnamefont
  {Sinayskiy}}\ and\ \bibinfo {author} {\bibfnamefont {F.}~\bibnamefont
  {Petruccione}},\ }\href@noop {} {\bibfield  {journal} {\bibinfo  {journal}
  {Quantum Information Processing}\ }\textbf {\bibinfo {volume} {11}},\
  \bibinfo {pages} {1301} (\bibinfo {year} {2012})}\BibitemShut {NoStop}%
\bibitem [{\citenamefont {Mochizuki}\ \emph {et~al.}(2016)\citenamefont
  {Mochizuki}, \citenamefont {Kim},\ and\ \citenamefont
  {Obuse}}]{mochizuki2016explicit}%
  \BibitemOpen
  \bibfield  {author} {\bibinfo {author} {\bibfnamefont {K.}~\bibnamefont
  {Mochizuki}}, \bibinfo {author} {\bibfnamefont {D.}~\bibnamefont {Kim}}, \
  and\ \bibinfo {author} {\bibfnamefont {H.}~\bibnamefont {Obuse}},\
  }\href@noop {} {\bibfield  {journal} {\bibinfo  {journal} {Physical Review
  A}\ }\textbf {\bibinfo {volume} {93}},\ \bibinfo {pages} {062116} (\bibinfo
  {year} {2016})}\BibitemShut {NoStop}%
\bibitem [{\citenamefont {Mittal}\ \emph {et~al.}(2021)\citenamefont {Mittal},
  \citenamefont {Raj}, \citenamefont {Dey},\ and\ \citenamefont
  {Goyal}}]{mittal2021persistence}%
  \BibitemOpen
  \bibfield  {author} {\bibinfo {author} {\bibfnamefont {V.}~\bibnamefont
  {Mittal}}, \bibinfo {author} {\bibfnamefont {A.}~\bibnamefont {Raj}},
  \bibinfo {author} {\bibfnamefont {S.}~\bibnamefont {Dey}}, \ and\ \bibinfo
  {author} {\bibfnamefont {S.~K.}\ \bibnamefont {Goyal}},\ }\href@noop {}
  {\bibfield  {journal} {\bibinfo  {journal} {Scientific Reports}\ }\textbf
  {\bibinfo {volume} {11}},\ \bibinfo {pages} {1} (\bibinfo {year}
  {2021})}\BibitemShut {NoStop}%
\bibitem [{\citenamefont {Xiao}\ \emph {et~al.}(2017)\citenamefont {Xiao},
  \citenamefont {Zhan}, \citenamefont {Bian}, \citenamefont {Wang},
  \citenamefont {Zhang}, \citenamefont {Wang}, \citenamefont {Li},
  \citenamefont {Mochizuki}, \citenamefont {Kim}, \citenamefont {Kawakami}
  \emph {et~al.}}]{xiao2017observation}%
  \BibitemOpen
  \bibfield  {author} {\bibinfo {author} {\bibfnamefont {L.}~\bibnamefont
  {Xiao}}, \bibinfo {author} {\bibfnamefont {X.}~\bibnamefont {Zhan}}, \bibinfo
  {author} {\bibfnamefont {Z.}~\bibnamefont {Bian}}, \bibinfo {author}
  {\bibfnamefont {K.}~\bibnamefont {Wang}}, \bibinfo {author} {\bibfnamefont
  {X.}~\bibnamefont {Zhang}}, \bibinfo {author} {\bibfnamefont
  {X.}~\bibnamefont {Wang}}, \bibinfo {author} {\bibfnamefont {J.}~\bibnamefont
  {Li}}, \bibinfo {author} {\bibfnamefont {K.}~\bibnamefont {Mochizuki}},
  \bibinfo {author} {\bibfnamefont {D.}~\bibnamefont {Kim}}, \bibinfo {author}
  {\bibfnamefont {N.}~\bibnamefont {Kawakami}},  \emph {et~al.},\ }\href@noop
  {} {\bibfield  {journal} {\bibinfo  {journal} {Nature Physics}\ }\textbf
  {\bibinfo {volume} {13}},\ \bibinfo {pages} {1117} (\bibinfo {year}
  {2017})}\BibitemShut {NoStop}%
\bibitem [{\citenamefont {Dadras}\ \emph {et~al.}(2018)\citenamefont {Dadras},
  \citenamefont {Gresch}, \citenamefont {Groiseau}, \citenamefont {Wimberger},\
  and\ \citenamefont {Summy}}]{dadras2018quantum}%
  \BibitemOpen
  \bibfield  {author} {\bibinfo {author} {\bibfnamefont {S.}~\bibnamefont
  {Dadras}}, \bibinfo {author} {\bibfnamefont {A.}~\bibnamefont {Gresch}},
  \bibinfo {author} {\bibfnamefont {C.}~\bibnamefont {Groiseau}}, \bibinfo
  {author} {\bibfnamefont {S.}~\bibnamefont {Wimberger}}, \ and\ \bibinfo
  {author} {\bibfnamefont {G.~S.}\ \bibnamefont {Summy}},\ }\href@noop {}
  {\bibfield  {journal} {\bibinfo  {journal} {Physical review letters}\
  }\textbf {\bibinfo {volume} {121}},\ \bibinfo {pages} {070402} (\bibinfo
  {year} {2018})}\BibitemShut {NoStop}%
\bibitem [{\citenamefont {Ryu}\ and\ \citenamefont
  {Hatsugai}(2002)}]{ryu2002topological}%
  \BibitemOpen
  \bibfield  {author} {\bibinfo {author} {\bibfnamefont {S.}~\bibnamefont
  {Ryu}}\ and\ \bibinfo {author} {\bibfnamefont {Y.}~\bibnamefont {Hatsugai}},\
  }\href@noop {} {\bibfield  {journal} {\bibinfo  {journal} {Physical review
  letters}\ }\textbf {\bibinfo {volume} {89}},\ \bibinfo {pages} {077002}
  (\bibinfo {year} {2002})}\BibitemShut {NoStop}%
\bibitem [{\citenamefont {Bender}\ and\ \citenamefont
  {Boettcher}(1998)}]{bender1998real}%
  \BibitemOpen
  \bibfield  {author} {\bibinfo {author} {\bibfnamefont {C.~M.}\ \bibnamefont
  {Bender}}\ and\ \bibinfo {author} {\bibfnamefont {S.}~\bibnamefont
  {Boettcher}},\ }\href@noop {} {\bibfield  {journal} {\bibinfo  {journal}
  {Physical Review Letters}\ }\textbf {\bibinfo {volume} {80}},\ \bibinfo
  {pages} {5243} (\bibinfo {year} {1998})}\BibitemShut {NoStop}%
\bibitem [{\citenamefont {Bender}\ \emph {et~al.}(2002)\citenamefont {Bender},
  \citenamefont {Brody},\ and\ \citenamefont {Jones}}]{bender2002complex}%
  \BibitemOpen
  \bibfield  {author} {\bibinfo {author} {\bibfnamefont {C.~M.}\ \bibnamefont
  {Bender}}, \bibinfo {author} {\bibfnamefont {D.~C.}\ \bibnamefont {Brody}}, \
  and\ \bibinfo {author} {\bibfnamefont {H.~F.}\ \bibnamefont {Jones}},\
  }\href@noop {} {\bibfield  {journal} {\bibinfo  {journal} {Physical Review
  Letters}\ }\textbf {\bibinfo {volume} {89}},\ \bibinfo {pages} {270401}
  (\bibinfo {year} {2002})}\BibitemShut {NoStop}%
\bibitem [{\citenamefont {Yao}\ and\ \citenamefont {Wang}(2018)}]{yao2018edge}%
  \BibitemOpen
  \bibfield  {author} {\bibinfo {author} {\bibfnamefont {S.}~\bibnamefont
  {Yao}}\ and\ \bibinfo {author} {\bibfnamefont {Z.}~\bibnamefont {Wang}},\
  }\href@noop {} {\bibfield  {journal} {\bibinfo  {journal} {Physical review
  letters}\ }\textbf {\bibinfo {volume} {121}},\ \bibinfo {pages} {086803}
  (\bibinfo {year} {2018})}\BibitemShut {NoStop}%
\bibitem [{\citenamefont {Kunst}\ \emph {et~al.}(2018)\citenamefont {Kunst},
  \citenamefont {Edvardsson}, \citenamefont {Budich},\ and\ \citenamefont
  {Bergholtz}}]{kunst2018biorthogonal}%
  \BibitemOpen
  \bibfield  {author} {\bibinfo {author} {\bibfnamefont {F.~K.}\ \bibnamefont
  {Kunst}}, \bibinfo {author} {\bibfnamefont {E.}~\bibnamefont {Edvardsson}},
  \bibinfo {author} {\bibfnamefont {J.~C.}\ \bibnamefont {Budich}}, \ and\
  \bibinfo {author} {\bibfnamefont {E.~J.}\ \bibnamefont {Bergholtz}},\
  }\href@noop {} {\bibfield  {journal} {\bibinfo  {journal} {Physical review
  letters}\ }\textbf {\bibinfo {volume} {121}},\ \bibinfo {pages} {026808}
  (\bibinfo {year} {2018})}\BibitemShut {NoStop}%
\bibitem [{\citenamefont {Mochizuki}\ \emph {et~al.}(2020)\citenamefont
  {Mochizuki}, \citenamefont {Kim}, \citenamefont {Kawakami},\ and\
  \citenamefont {Obuse}}]{mochizuki2020bulk}%
  \BibitemOpen
  \bibfield  {author} {\bibinfo {author} {\bibfnamefont {K.}~\bibnamefont
  {Mochizuki}}, \bibinfo {author} {\bibfnamefont {D.}~\bibnamefont {Kim}},
  \bibinfo {author} {\bibfnamefont {N.}~\bibnamefont {Kawakami}}, \ and\
  \bibinfo {author} {\bibfnamefont {H.}~\bibnamefont {Obuse}},\ }\href@noop {}
  {\bibfield  {journal} {\bibinfo  {journal} {Physical Review A}\ }\textbf
  {\bibinfo {volume} {102}},\ \bibinfo {pages} {062202} (\bibinfo {year}
  {2020})}\BibitemShut {NoStop}%
\bibitem [{\citenamefont {Wei{\ss}enhofer}\ and\ \citenamefont
  {Nowak}(2020)}]{weissenhofer2020diffusion}%
  \BibitemOpen
  \bibfield  {author} {\bibinfo {author} {\bibfnamefont {M.}~\bibnamefont
  {Wei{\ss}enhofer}}\ and\ \bibinfo {author} {\bibfnamefont {U.}~\bibnamefont
  {Nowak}},\ }\href@noop {} {\bibfield  {journal} {\bibinfo  {journal} {New
  Journal of Physics}\ }\textbf {\bibinfo {volume} {22}},\ \bibinfo {pages}
  {103059} (\bibinfo {year} {2020})}\BibitemShut {NoStop}%
\bibitem [{\citenamefont {Yoshida}\ and\ \citenamefont
  {Hatsugai}(2021)}]{yoshida2021bulk}%
  \BibitemOpen
  \bibfield  {author} {\bibinfo {author} {\bibfnamefont {T.}~\bibnamefont
  {Yoshida}}\ and\ \bibinfo {author} {\bibfnamefont {Y.}~\bibnamefont
  {Hatsugai}},\ }\href@noop {} {\bibfield  {journal} {\bibinfo  {journal}
  {Scientific reports}\ }\textbf {\bibinfo {volume} {11}},\ \bibinfo {pages}
  {1} (\bibinfo {year} {2021})}\BibitemShut {NoStop}%
\bibitem [{\citenamefont {Zeuner}\ \emph {et~al.}(2015)\citenamefont {Zeuner},
  \citenamefont {Rechtsman}, \citenamefont {Plotnik}, \citenamefont {Lumer},
  \citenamefont {Nolte}, \citenamefont {Rudner}, \citenamefont {Segev},\ and\
  \citenamefont {Szameit}}]{zeuner2015observation}%
  \BibitemOpen
  \bibfield  {author} {\bibinfo {author} {\bibfnamefont {J.~M.}\ \bibnamefont
  {Zeuner}}, \bibinfo {author} {\bibfnamefont {M.~C.}\ \bibnamefont
  {Rechtsman}}, \bibinfo {author} {\bibfnamefont {Y.}~\bibnamefont {Plotnik}},
  \bibinfo {author} {\bibfnamefont {Y.}~\bibnamefont {Lumer}}, \bibinfo
  {author} {\bibfnamefont {S.}~\bibnamefont {Nolte}}, \bibinfo {author}
  {\bibfnamefont {M.~S.}\ \bibnamefont {Rudner}}, \bibinfo {author}
  {\bibfnamefont {M.}~\bibnamefont {Segev}}, \ and\ \bibinfo {author}
  {\bibfnamefont {A.}~\bibnamefont {Szameit}},\ }\href@noop {} {\bibfield
  {journal} {\bibinfo  {journal} {Physical review letters}\ }\textbf {\bibinfo
  {volume} {115}},\ \bibinfo {pages} {040402} (\bibinfo {year}
  {2015})}\BibitemShut {NoStop}%
\bibitem [{\citenamefont {Weimann}\ \emph {et~al.}(2017)\citenamefont
  {Weimann}, \citenamefont {Kremer}, \citenamefont {Plotnik}, \citenamefont
  {Lumer}, \citenamefont {Nolte}, \citenamefont {Makris}, \citenamefont
  {Segev}, \citenamefont {Rechtsman},\ and\ \citenamefont
  {Szameit}}]{weimann2017topologically}%
  \BibitemOpen
  \bibfield  {author} {\bibinfo {author} {\bibfnamefont {S.}~\bibnamefont
  {Weimann}}, \bibinfo {author} {\bibfnamefont {M.}~\bibnamefont {Kremer}},
  \bibinfo {author} {\bibfnamefont {Y.}~\bibnamefont {Plotnik}}, \bibinfo
  {author} {\bibfnamefont {Y.}~\bibnamefont {Lumer}}, \bibinfo {author}
  {\bibfnamefont {S.}~\bibnamefont {Nolte}}, \bibinfo {author} {\bibfnamefont
  {K.~G.}\ \bibnamefont {Makris}}, \bibinfo {author} {\bibfnamefont
  {M.}~\bibnamefont {Segev}}, \bibinfo {author} {\bibfnamefont {M.~C.}\
  \bibnamefont {Rechtsman}}, \ and\ \bibinfo {author} {\bibfnamefont
  {A.}~\bibnamefont {Szameit}},\ }\href@noop {} {\bibfield  {journal} {\bibinfo
   {journal} {Nature materials}\ }\textbf {\bibinfo {volume} {16}},\ \bibinfo
  {pages} {433} (\bibinfo {year} {2017})}\BibitemShut {NoStop}%
\bibitem [{\citenamefont {Essin}\ and\ \citenamefont
  {Gurarie}(2011)}]{essin2011bulk}%
  \BibitemOpen
  \bibfield  {author} {\bibinfo {author} {\bibfnamefont {A.~M.}\ \bibnamefont
  {Essin}}\ and\ \bibinfo {author} {\bibfnamefont {V.}~\bibnamefont
  {Gurarie}},\ }\href@noop {} {\bibfield  {journal} {\bibinfo  {journal}
  {Physical Review B}\ }\textbf {\bibinfo {volume} {84}},\ \bibinfo {pages}
  {125132} (\bibinfo {year} {2011})}\BibitemShut {NoStop}%
\bibitem [{\citenamefont {Brun}\ \emph {et~al.}(2003)\citenamefont {Brun},
  \citenamefont {Carteret},\ and\ \citenamefont {Ambainis}}]{brun2003quantum}%
  \BibitemOpen
  \bibfield  {author} {\bibinfo {author} {\bibfnamefont {T.~A.}\ \bibnamefont
  {Brun}}, \bibinfo {author} {\bibfnamefont {H.~A.}\ \bibnamefont {Carteret}},
  \ and\ \bibinfo {author} {\bibfnamefont {A.}~\bibnamefont {Ambainis}},\
  }\href@noop {} {\bibfield  {journal} {\bibinfo  {journal} {Physical review
  letters}\ }\textbf {\bibinfo {volume} {91}},\ \bibinfo {pages} {130602}
  (\bibinfo {year} {2003})}\BibitemShut {NoStop}%
\bibitem [{\citenamefont {Romanelli}\ \emph {et~al.}(2005)\citenamefont
  {Romanelli}, \citenamefont {Siri}, \citenamefont {Abal}, \citenamefont
  {Auyuanet},\ and\ \citenamefont {Donangelo}}]{romanelli2005decoherence}%
  \BibitemOpen
  \bibfield  {author} {\bibinfo {author} {\bibfnamefont {A.}~\bibnamefont
  {Romanelli}}, \bibinfo {author} {\bibfnamefont {R.}~\bibnamefont {Siri}},
  \bibinfo {author} {\bibfnamefont {G.}~\bibnamefont {Abal}}, \bibinfo {author}
  {\bibfnamefont {A.}~\bibnamefont {Auyuanet}}, \ and\ \bibinfo {author}
  {\bibfnamefont {R.}~\bibnamefont {Donangelo}},\ }\href@noop {} {\bibfield
  {journal} {\bibinfo  {journal} {Physica A: Statistical Mechanics and its
  Applications}\ }\textbf {\bibinfo {volume} {347}},\ \bibinfo {pages} {137}
  (\bibinfo {year} {2005})}\BibitemShut {NoStop}%
\bibitem [{\citenamefont {Xue}\ \emph {et~al.}(2014)\citenamefont {Xue},
  \citenamefont {Qin},\ and\ \citenamefont {Tang}}]{xue2014trapping}%
  \BibitemOpen
  \bibfield  {author} {\bibinfo {author} {\bibfnamefont {P.}~\bibnamefont
  {Xue}}, \bibinfo {author} {\bibfnamefont {H.}~\bibnamefont {Qin}}, \ and\
  \bibinfo {author} {\bibfnamefont {B.}~\bibnamefont {Tang}},\ }\href@noop {}
  {\bibfield  {journal} {\bibinfo  {journal} {Scientific reports}\ }\textbf
  {\bibinfo {volume} {4}},\ \bibinfo {pages} {1} (\bibinfo {year}
  {2014})}\BibitemShut {NoStop}%
\bibitem [{\citenamefont {Cardano}\ \emph {et~al.}(2016)\citenamefont
  {Cardano}, \citenamefont {Maffei}, \citenamefont {Massa}, \citenamefont
  {Piccirillo}, \citenamefont {De~Lisio}, \citenamefont {De~Filippis},
  \citenamefont {Cataudella}, \citenamefont {Santamato},\ and\ \citenamefont
  {Marrucci}}]{cardano2016statistical}%
  \BibitemOpen
  \bibfield  {author} {\bibinfo {author} {\bibfnamefont {F.}~\bibnamefont
  {Cardano}}, \bibinfo {author} {\bibfnamefont {M.}~\bibnamefont {Maffei}},
  \bibinfo {author} {\bibfnamefont {F.}~\bibnamefont {Massa}}, \bibinfo
  {author} {\bibfnamefont {B.}~\bibnamefont {Piccirillo}}, \bibinfo {author}
  {\bibfnamefont {C.}~\bibnamefont {De~Lisio}}, \bibinfo {author}
  {\bibfnamefont {G.}~\bibnamefont {De~Filippis}}, \bibinfo {author}
  {\bibfnamefont {V.}~\bibnamefont {Cataudella}}, \bibinfo {author}
  {\bibfnamefont {E.}~\bibnamefont {Santamato}}, \ and\ \bibinfo {author}
  {\bibfnamefont {L.}~\bibnamefont {Marrucci}},\ }\href@noop {} {\bibfield
  {journal} {\bibinfo  {journal} {Nature communications}\ }\textbf {\bibinfo
  {volume} {7}},\ \bibinfo {pages} {1} (\bibinfo {year} {2016})}\BibitemShut
  {NoStop}%
\bibitem [{\citenamefont {Wang}\ \emph {et~al.}(2018)\citenamefont {Wang},
  \citenamefont {Xiao}, \citenamefont {Qiu}, \citenamefont {Wang},
  \citenamefont {Yi},\ and\ \citenamefont {Xue}}]{wang2018detecting}%
  \BibitemOpen
  \bibfield  {author} {\bibinfo {author} {\bibfnamefont {X.}~\bibnamefont
  {Wang}}, \bibinfo {author} {\bibfnamefont {L.}~\bibnamefont {Xiao}}, \bibinfo
  {author} {\bibfnamefont {X.}~\bibnamefont {Qiu}}, \bibinfo {author}
  {\bibfnamefont {K.}~\bibnamefont {Wang}}, \bibinfo {author} {\bibfnamefont
  {W.}~\bibnamefont {Yi}}, \ and\ \bibinfo {author} {\bibfnamefont
  {P.}~\bibnamefont {Xue}},\ }\href@noop {} {\bibfield  {journal} {\bibinfo
  {journal} {Physical Review A}\ }\textbf {\bibinfo {volume} {98}},\ \bibinfo
  {pages} {013835} (\bibinfo {year} {2018})}\BibitemShut {NoStop}%
\bibitem [{\citenamefont {Xiao}\ \emph {et~al.}(2018)\citenamefont {Xiao},
  \citenamefont {Qiu}, \citenamefont {Wang}, \citenamefont {Bian},
  \citenamefont {Zhan}, \citenamefont {Obuse}, \citenamefont {Sanders},
  \citenamefont {Yi},\ and\ \citenamefont {Xue}}]{xiao2018higher}%
  \BibitemOpen
  \bibfield  {author} {\bibinfo {author} {\bibfnamefont {L.}~\bibnamefont
  {Xiao}}, \bibinfo {author} {\bibfnamefont {X.}~\bibnamefont {Qiu}}, \bibinfo
  {author} {\bibfnamefont {K.}~\bibnamefont {Wang}}, \bibinfo {author}
  {\bibfnamefont {Z.}~\bibnamefont {Bian}}, \bibinfo {author} {\bibfnamefont
  {X.}~\bibnamefont {Zhan}}, \bibinfo {author} {\bibfnamefont {H.}~\bibnamefont
  {Obuse}}, \bibinfo {author} {\bibfnamefont {B.~C.}\ \bibnamefont {Sanders}},
  \bibinfo {author} {\bibfnamefont {W.}~\bibnamefont {Yi}}, \ and\ \bibinfo
  {author} {\bibfnamefont {P.}~\bibnamefont {Xue}},\ }\href@noop {} {\bibfield
  {journal} {\bibinfo  {journal} {Physical Review A}\ }\textbf {\bibinfo
  {volume} {98}},\ \bibinfo {pages} {063847} (\bibinfo {year}
  {2018})}\BibitemShut {NoStop}%
\bibitem [{\citenamefont {Xiaoxia}\ and\ \citenamefont
  {Zhijian}(2017)}]{zhang2017topology}%
  \BibitemOpen
  \bibfield  {author} {\bibinfo {author} {\bibfnamefont {Z.}~\bibnamefont
  {Xiaoxia}}\ and\ \bibinfo {author} {\bibfnamefont {L.}~\bibnamefont
  {Zhijian}},\ }\href@noop {} {\bibfield  {journal} {\bibinfo  {journal}
  {Journal of Shanxi University}\ }\textbf {\bibinfo {volume} {40}},\ \bibinfo
  {pages} {100} (\bibinfo {year} {2017})}\BibitemShut {NoStop}%
\bibitem [{\citenamefont {L{\'o}pez}\ and\ \citenamefont
  {Paz}(2003)}]{lopez2003phase}%
  \BibitemOpen
  \bibfield  {author} {\bibinfo {author} {\bibfnamefont {C.~C.}\ \bibnamefont
  {L{\'o}pez}}\ and\ \bibinfo {author} {\bibfnamefont {J.~P.}\ \bibnamefont
  {Paz}},\ }\href@noop {} {\bibfield  {journal} {\bibinfo  {journal} {Physical
  Review A}\ }\textbf {\bibinfo {volume} {68}},\ \bibinfo {pages} {052305}
  (\bibinfo {year} {2003})}\BibitemShut {NoStop}%
\bibitem [{\citenamefont {Kendon}\ and\ \citenamefont
  {Tregenna}(2003)}]{kendon2003decoherence}%
  \BibitemOpen
  \bibfield  {author} {\bibinfo {author} {\bibfnamefont {V.}~\bibnamefont
  {Kendon}}\ and\ \bibinfo {author} {\bibfnamefont {B.}~\bibnamefont
  {Tregenna}},\ }\href@noop {} {\bibfield  {journal} {\bibinfo  {journal}
  {Physical Review A}\ }\textbf {\bibinfo {volume} {67}},\ \bibinfo {pages}
  {042315} (\bibinfo {year} {2003})}\BibitemShut {NoStop}%
\end{thebibliography}%

\end{document}